\documentclass[runningheads]{llncs}
\usepackage{xcolor}
\usepackage{amssymb}
\usepackage{amsmath}
\usepackage[T1]{fontenc}
\usepackage{graphicx,verbatim}
\usepackage{color}
\usepackage{multirow}
\begin{document}
\title{Pathology Foundation Models are Scanner Sensitive: Benchmark and Mitigation with Contrastive \textit{ScanGen} Loss}
\definecolor{mygreen}{rgb}{0.1,0.4,0.1}
\definecolor{myred}{rgb}{0.4,0.1,0.1}
\newcommand{\repeatthanks}{\textsuperscript{\thefootnote}}
\author{Gianluca Carloni\inst{1}\thanks{Shared first authorship}\orcidID{0000-0002-5774-361X} \and
Biagio Brattoli\inst{1}\repeatthanks\orcidID{0000-0002-8482-3375} \and
Seongho Keum\inst{1}\orcidID{0009-0008-0954-9389} \and
Jongchan Park\inst{1}\orcidID{0000-0001-9808-6823} \and
Taebum Lee\inst{1}\orcidID{0000-0002-9472-0644} \and
Chang Ho Ahn\inst{1}\orcidID{0000-0002-0702-0608} \and
Sergio Pereira\inst{1}\orcidID{0000-0002-4298-0903}
}
\authorrunning{G. Carloni, B. Brattoli et al.}
%
\institute{Lunit, Seoul, Republic of Korea \\
    \email{gianluca.carloni@lunit.io} \\
}

\titlerunning{Pathology FMs are Scanner Sensitive: Benchmark \& Mitigation}
\maketitle              
\begin{abstract}
Computational pathology (CPath) has shown great potential in mining actionable insights from Whole Slide Images (WSIs). Deep Learning (DL) has been at the center of modern CPath, and while it delivers unprecedented performance, it is also known that DL may be affected by irrelevant details, such as those introduced during scanning by different commercially available scanners. This may lead to \textit{scanner bias}, where the model outputs for the same tissue acquired by different scanners may vary. In turn, it hinders the trust of clinicians in CPath-based tools and their deployment in real-world clinical practices. Recent pathology Foundation Models (FMs) promise to provide better domain generalization capabilities. In this paper, we benchmark FMs using a multi-scanner dataset and show that FMs still suffer from scanner bias. Following this observation, we propose \textit{ScanGen}, a contrastive loss function applied during task-specific fine-tuning that mitigates scanner bias, thereby enhancing the models' robustness to scanner variations. Our approach is applied to the Multiple Instance Learning task of Epidermal Growth Factor Receptor (EGFR) mutation prediction from H\&E-stained WSIs in lung cancer. We observe that \textit{ScanGen} notably enhances the ability to generalize across scanners, while retaining or improving the performance of EGFR mutation prediction.


\end{abstract}

\section{Introduction}

\begin{figure}[t!]
\includegraphics[width=\textwidth]{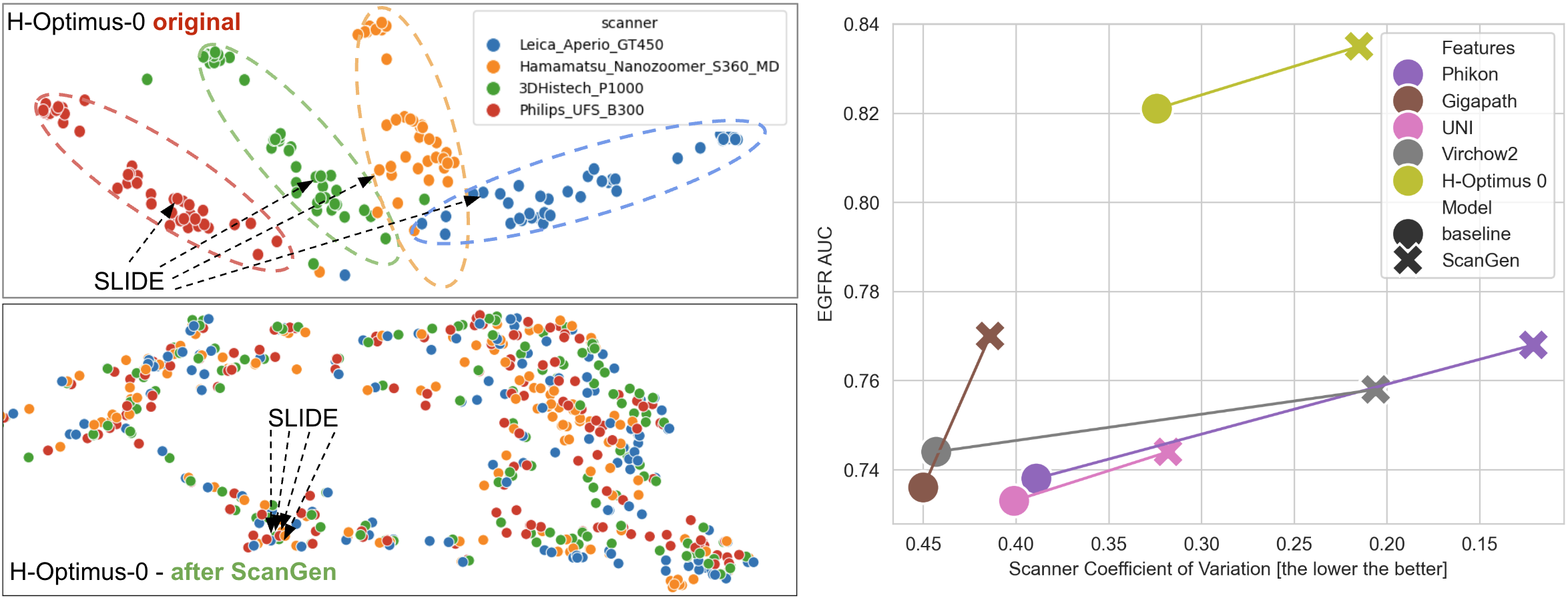}
\caption{Left: Existing pathology Foundation Models suffer from scanner bias. We show a 2D projection of WSIs (dots) from 4 scanners (colors) using UMAP based on H-Optimus-0 embeddings before and after \textit{ScanGen}. "SLIDE$\xrightarrow{}$" indicates the slides of the same specimen from different scanners. Right: \textit{ScanGen} improves both generalization and performance.}
\label{fig:intro}
\end{figure}

Digital pathology is catalyzing a new era in the field of histopathology by transforming how clinicians interact with and analyze tissue samples. Furthermore, it enables computational pathology (CPath) over whole slide images (WSIs), a key component for deriving novel and quantitative insights from WSIs. This brings new capabilities to clinical practice, such as prediction of genetic alterations \cite{kather2020pan,teichmann2022end}, tumor micro-environment analysis \cite{lunit2022jco}, or survival prediction \cite{ramanathan2024ensemble}.

Digital pathology requires the digitization of tissue slides into WSIs using slide scanners. While these scanners must capture the underlying tissue accurately, different devices may result in differences in image characteristics \cite{duenweg2023whole}. For example, differences in optical systems may impact image sharpness, variations in light sources can affect contrast and introduce artifacts, and proprietary software (e.g. for data compression) can affect image processing. To implement CPath-based tools in clinical settings, it is crucial to address the constraints specific to different scanners. Since different scanners are used worldwide, with some being predominant in certain regions (e.g., Hamamatsu in Asia \cite{pinto2023exploring}), any performance issues associated with a particular scanner could lead to regional disparities in patient care.

Recently, the introduction of foundation models (FMs) provides a base upon which CPath tools are developed. However, while FMs promise better generalization capabilities, they may still encode several biases, such as the slide identity \cite{exaonepath} and hospital environments \cite{unrobustfm}.
The present paper delves into another crucial aspect of CPath, the digital scanners used during image acquisition, offering empirical evidence that FMs are sensitive to the type of digital scanner employed for reading tissue specimens.
Since CPath FMs are typically used as frozen extractors with no access to the input images, alleviating this bias with traditional stain normalization methods (e.g., Reinhard\cite{reinhard2001color}, Macenko \cite{macenko2009method}) or DL ones (e.g., StainGAN\cite{shaban2019staingan}, StainNet\cite{kang2021stainnet}) is not trivial.
Scanner bias results in weakened generalization capabilities for FM-based CPath models, posing challenges in their broader application. In particular, scanner bias means that the model output will change based on which scanner has been used to read the specimen despite the content being the same, which can significantly hinder trust in CPath-based tools and, consequently, slower clinical adoption.

To address this shortcoming, we propose \textit{ScanGen}, a contrastive loss function that mitigates scanner bias, thereby enhancing the model's robustness across different scanners. Our evaluation involves testing various state-of-the-art FMs using a benchmark set that comprises images of the same specimen captured from six commercially available scanners at 40$\times$ magnification and one scanner at 20$\times$ magnification. Additionally, our approach is applied to the task of detecting Epidermal Growth Factor Receptor (EGFR) mutations in lung cancer, a prevalent and clinically significant endeavor in the CPath space \cite{campanella2022h,zhao2023high}. 

In summary, this paper introduces three contributions: 1) a contrastive loss function aimed at reducing scanner bias in Multiple Instance Learning (MIL), 2) an assessment of FMs in relation to scanner generalization, and 3) application of the proposed method to EGFR mutation prediction in lung cancer, a clinically relevant task.
The results demonstrate promising improvements in predictive performance and generalization (Fig. \ref{fig:intro}), underscoring the potential of our methodologies in clinical settings.

\section{Method}

\begin{figure}[t!]
\includegraphics[width=\textwidth]{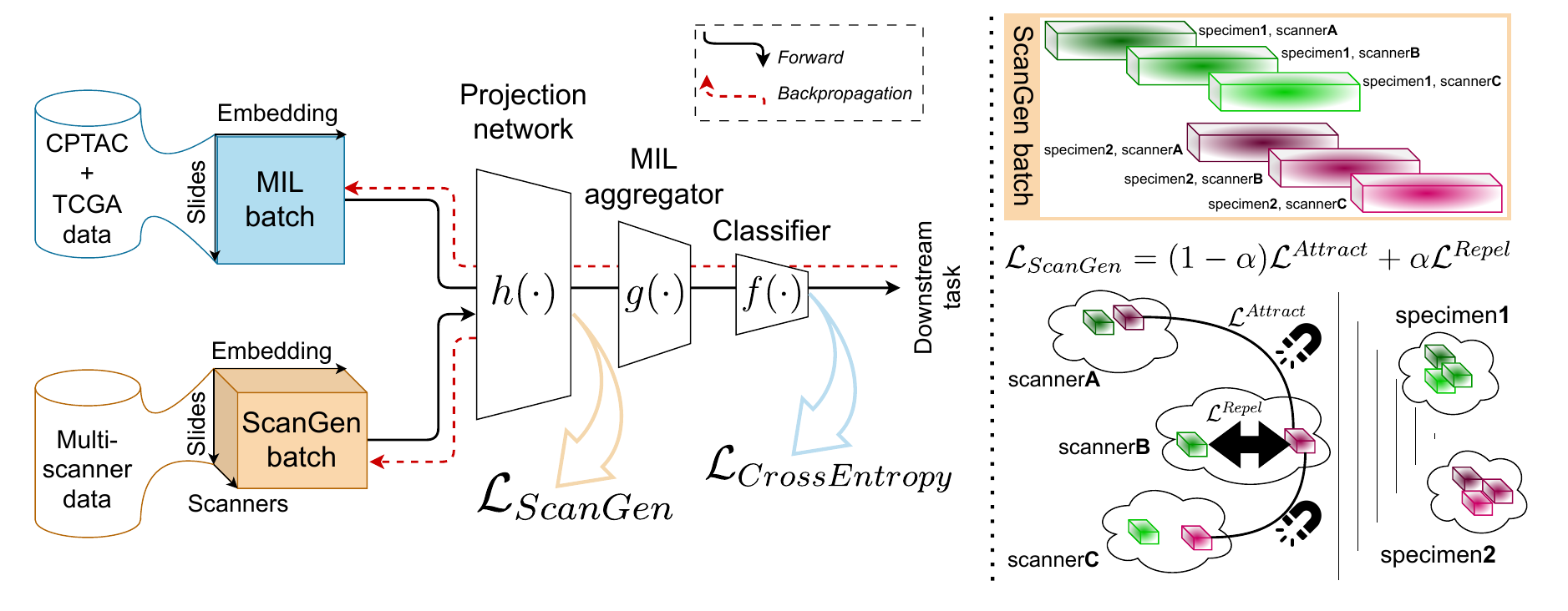}
\caption{Our method. Left: Data are projected to a new embedding space where scanner bias is removed via our contrastive \textit{ScanGen} loss. \textit{ScanGen} is trained on paired WSIs of the same specimen acquired with different scanners and can be easily integrated with any existing method by acting before the MIL aggregator. Right: \textit{ScanGen} attracts the representations of same-specimen different-scanner pairs, while repelling same-scanner different-specimen ones. This favors specimen-based clusters to scanner-based ones, thus fostering generalization across scanners in downstream applications.}
\label{fig:method}
\end{figure}

\subsection{Scanner Generalization Metric for MIL}
Due to the extremely large size of WSIs, MIL is usually employed for classification or regression tasks. 
In MIL, WSIs are partitioned into $P$ patches and the $FM(\cdot)$ is used as a pretrained, frozen feature extractor to obtain an embedding vector $x_i = FM(p_i)$ per each patch $p_i$. Finally, an aggregator function $g(\cdot)$ combines all patches and a classifier $f(\cdot)$ produces the final prediction (e.g., EGFR mutation) as $\hat{y} = f(g(x_1,\dots,x_P))$.
In this context, the scanner bias relates to how different two predictions $\hat{y}^{s_1}_m$ and $\hat{y}^{s_2}_m$ are from the same specimen $m$ but different scanners $s_1$ and $s_2$.
For this reason, we measure the intensity of scanner bias as the average coefficient of variation (CoV) between predictions of the same specimen $m$ across $S$ multiple scanners, defined as
\begin{equation}
\mathrm{CoV} = \frac{1}{M} \sum_{m=1}^{M} \left( \frac{\sigma\bigl(\hat{y}_m^{s_1},\hat{y}_m^{s_2},\dots,\hat{y}_m^{s_S}\bigr)}{\left|\mu\bigl(\hat{y}_m^{s_1},\hat{y}_m^{s_2},\dots,\hat{y}_m^{s_S}\bigr)\right|} \right),
\end{equation}

\noindent with $M$ number of specimens, where $\sigma$ and $\mu$ are, respectively, the standard deviation and average across the logit predictions of the same specimen acquired with different scanner hardware. 
CoV is more appropriate than simple standard deviation because it allows for meaningful comparisons across sources with different scales.
Lower $\mathrm{CoV}$ means better agreement in predictions, indicating the model is less scanner-sensitive.

\subsection{Proposed Framework and \textit{ScanGen} Loss}

As Figure \ref{fig:method} illustrates, to reduce the scanner bias in the embedding space, we first map the patch embeddings $x$ to a new scanner-unbiased embedding space using a projection network $h(\cdot)$, which is trained as follows. 
Given two specimens $m$ and $n$ and two scanners $s_1$ and $s_2$, we rework the contrastive loss function\cite{hadsell2006dimensionality} and define our \textit{ScanGen} loss as
\begin{equation}
\label{eq:scangenloss}
\mathcal{L}_{ScanGen} = (1-\alpha) \underbrace{\left(d(h(x_m^{s_1}),h(x_m^{s_2}))\right)^2}_\text{$\mathcal{L}^{Attract}$} + \alpha \underbrace{\left(\max(0, r - d(h(x_m^{s_1}),h(x_n^{s_1}))\right)^2}_\text{$\mathcal{L}^{Repel}$}
\end{equation}
\noindent where 
$d(a,b)=1-\frac{a \cdot b}{|a|\cdot|b|}$ is the cosine distance function, $\mathcal{L}^{Attract}$ is the attraction term encouraging samples from the same specimen but different scanners to be close in the embedding space, $\mathcal{L}^{Repel}$ is the repulsion term pushing samples from the same scanner but different specimens apart, $r$ is the radius, and $\alpha$ is a hyper-parameter governing the weighted combination of those terms. The MIL aggregator is then applied: $g(\cdot) = g(h(x_1),h(x_2),\dots,h(x_P))$. 

The final loss is composed of the downstream task loss, in our case EGFR prediction using standard cross-entropy (CE), and our \textit{ScanGen} loss:

\begin{equation}
\label{eq:totalloss}
\mathcal{L} = \frac{1}{|A|} \sum_{a \in A} \mathcal{L}_{\text{CE}}(\hat{y}_a, y_a) + \lambda \cdot \frac{1}{|B|} \sum_{\substack{m,n \in B \\ m \neq n}} \sum_{\substack{u,v \in S \\ u \neq v}} \mathcal{L}_{\text{ScanGen}}(x_m^{s_u}, x_m^{s_v}, x_n^{s_u}; h)
\end{equation}

\noindent where $A$ and $B$ are independent sets of labeled samples for the downstream task and the \textit{ScanGen} task, respectively, $S$ is the number of scanners, and $\lambda$ is a hyper-parameter to control the balance between those two terms.

\subsection{Implementation}
Within a shared model, two different data flows occur at each forward pass during training (see Fig. \ref{fig:method}-left). On the one hand, a regular MIL batch sampled from an EGFR-labeled data pool enters the projection-aggregation-classification cascade, thus producing the logits for standard CE loss computation.

On the other hand, a batch of embeddings identified by specimen and scanner names is fed into the projection network $h(\cdot)$. There, the \textit{ScanGen} loss first calculates a distance matrix using pairwise cosine similarity between embeddings and constructs matrices to identify, within the batch, pairs of samples from the same scanner or with the same specimen name. These matrices are used to compute the attraction and repulsion losses, which are weighted and combined on the basis of the $\alpha$ parameter. The final loss is averaged over valid pairs, ensuring that only relevant pairs (i.e., either same-specimen different-scanners or same-scanner different-specimen) contribute to the loss calculation. For fair comparison, also the baseline models where \textit{ScanGen} is not enabled utilize the projection network $h(\cdot)$, which is learned, together with $g(\cdot)$ and $f(\cdot)$, under the CE loss alone.

We developed our code in PyTorch. We implement $h(\cdot)$ as a multi-layer perceptron (MLP) with three layers and hidden dimension between $48$ and $96$ depending on the FM features. Regarding $g(\cdot)$, we use the global average pooling (GAP) as a simple MIL model, where patch-level features are averaged to produce one embedding per WSI: $X = \frac{1}{P} \sum_{i=1}^P x_i$, followed by an MLP classifier.
Moreover, Tab. \ref{tab:sota} shows an adaptation to some modern MIL aggregators, for which we followed their official implementation. In our experiments, we found that $\alpha$ of $0.12$ to $0.21$, $r$ of $0.90$ to $1.10$, and $\lambda$ of $0.5$ to $1.0$ are appropriate.

\section{Results}
\subsection{Datasets}


Two publicly available datasets, CPTAC \cite{CPTAC} and TCGA \cite{tcga_luad}), for a total of $1,144$ WSIs, were used to train the EGFR branch. Note that only non-small cell lung cancer specimens were selected.
Our multi-scanner dataset, on the other hand, is composed of $323$ specimens, each tested for EGFR mutation using PCR and NGS assay as well as scanned using six devices: Leica Aperio AT2, Leica Aperio GT450, Hamamatsu Nanozoomer S360MD, 3DHistech P1000, Philips UFS B300, Roche Ventana DP200. The specimens were scanned at 40$\times$ magnification. The specimens were also scanned with Leica Aperio GT450 at 20$\times$ magnification. In total, the multi-scanner dataset is composed of $323 * 7 = 2,261$ WSIs. Fig. \ref{fig:ablation}-right shows different scans of the same specimen using the six different scanners.
We split this dataset into three splits while preserving class balance and specimen-level split (i.e., WSIs of the same specimen were grouped). As a result, $54$ specimens were used to train the \textit{ScanGen} branch, $45$ specimens were used as validation set for tuning both EGFR branch and \textit{ScanGen} branch, and $224$ specimens were used as held-out test set for external evaluation of downstream task performance (EGFR AUC) and scanner agreement (CoV). 

\subsection{Evaluating Scanner Generalization}
\begin{figure}[t!]
\includegraphics[width=\textwidth]{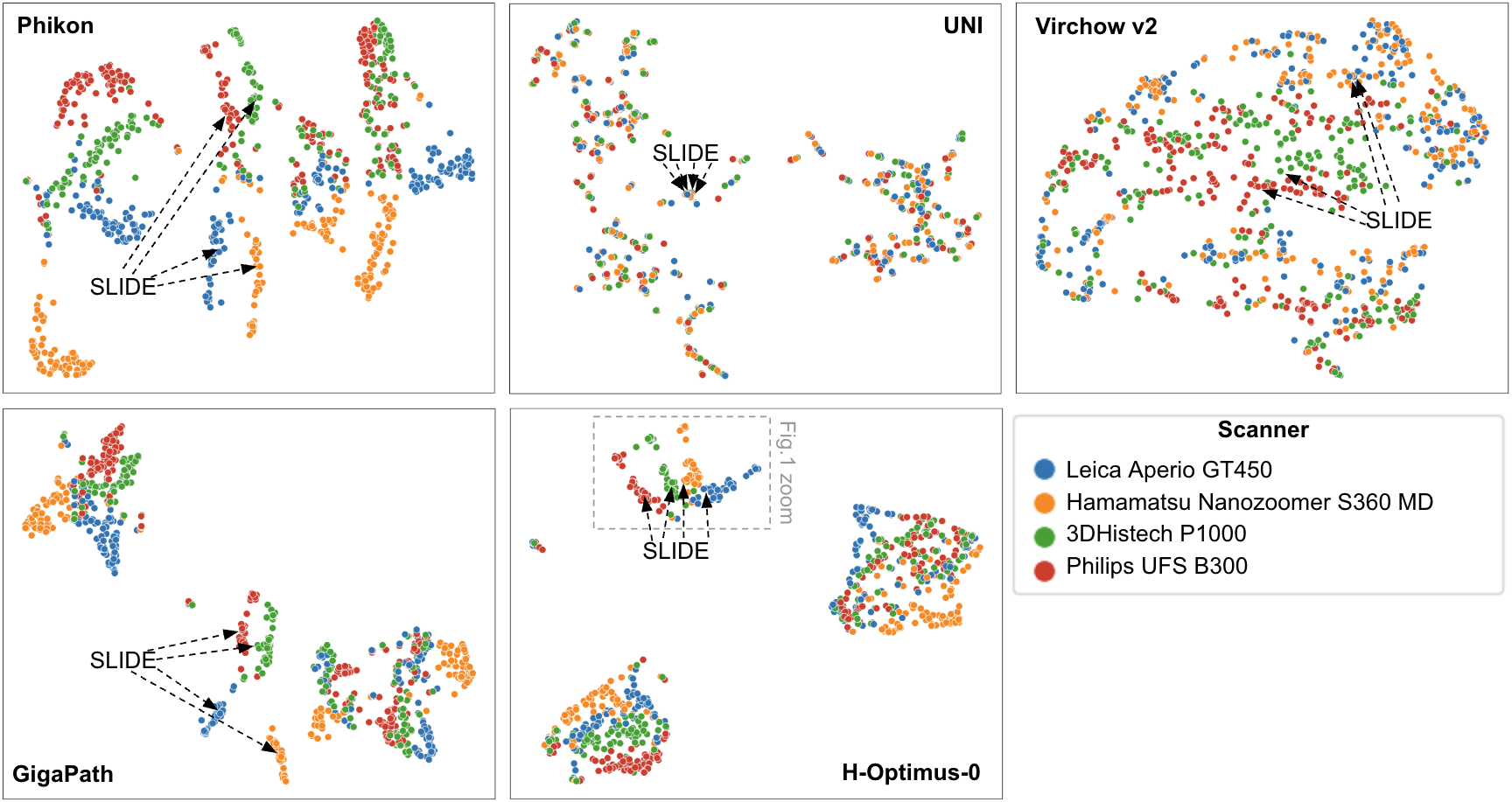}
\caption{Our multi-scanner dataset projected to 2D using various FMs. 
Each dot is a WSI embedding from the test set by averaging FM embeddings across patches and projecting them using UMAP. The color indicates the scanner. We selected 4 out of 7 scanners to reduce clutter. A single slide is indicated with "SLIDE$\xrightarrow{}$" to emphasize the location of the slides of the same specimen from different scanners in each plot.
}
\label{fig:tsne}
\end{figure}

We assess FMs on our multi-scanner dataset, presenting both qualitative and quantitative outcomes. 
To analyze the bias for each FM visually, for each WSI in the test set, we average all patch embeddings to produce a single WSI embedding, and project the full test set in 2D using UMAP \cite{mcinnes2018umap} with cosine metric and default parameters.
Fig. \ref{fig:tsne} illustrates that each FM is affected by scanner bias, as evidenced by the clustering and separation of samples according to the scanner rather than the same specimen. The only exception is UNI \cite{uni}, which, according to the 2D projection, seems to be unaffected by this bias; however, our quantitative results paint a different picture, and even UNI largely benefits from our \textit{ScanGen} loss.

In Tab. \ref{tab:scangen}, we benchmark five state-of-the-art FMs according to our new metrics for evaluating generalization, CoV over scanners and CoV over magnifications, and show \textit{ScanGen}'s mitigation effect. 
Our expectation is that \textit{ScanGen} would reduce the CoV, while avoiding hindrance on the EGFR downstream task.
Our findings indicate that incorporating \textit{ScanGen} projection significantly enhances the ability to generalize across scanners (lower CoV).
More importantly, the predictive performance for EGFR is not hindered; rather, it actually improves (higher AUC) for each embedding type.
Additionally, according to our findings, the \textit{ScanGen} loss also helps mitigate a disagreement in predictions when different magnifications are used to read the slide.

A similar result is shown in Fig. \ref{fig:intro}-right, where each model is compared on CoV and EGFR AUC, with and without \textit{ScanGen} loss. While CoV significantly improves for each FM, we also see a boost in EGFR AUC.

\begin{table}[t!]
\centering
\footnotesize
\caption{Evaluating scanner generalization for available foundation models on held-out testset. Coefficient of EGFR Variation (CoV) is shown across six scanners and two magnifications (40$\times$, 20$\times$).}\label{tab:scangen}
\begin{tabular}{|l|l|l|l|l|}
\hline
Embedding & Use \textit{ScanGen} & EGFR AUC [$\uparrow$] & Scanner CoV [$\downarrow$] & Magnification CoV [$\downarrow$] \\
\hline
Phikon\cite{phikon} & No & 0.738 & 0.389 & 0.158\\
 & Yes  & \textbf{0.768} & \textbf{0.121} \textcolor{mygreen}{(-68.9\%)} & \textbf{0.053} \textcolor{mygreen}{(-66.4\%)} \\
\hline
UNI\cite{uni} & No & 0.733 & 0.401 & 0.116\\ 
 & Yes  & \textbf{0.744} & \textbf{0.318} \textcolor{mygreen}{(-20.7\%)} & \textbf{0.089} \textcolor{mygreen}{(-23.3\%)} \\ 
\hline
Virchow\cite{virchow2} & No & 0.744 & 0.443 & 0.125 \\ 
 & Yes  & \textbf{0.758} & \textbf{0.206} \textcolor{mygreen}{(-53.5\%)} & \textbf{0.077} \textcolor{mygreen}{(-38.4\%)}\\ 
\hline
Gigapath\cite{gigapath} & No & 0.736 & 0.450 & \textbf{0.174} \\ 
 & Yes & \textbf{0.770} & \textbf{0.414} \textcolor{mygreen}{(-8.00\%)} & 0.193 \textcolor{myred}{(+10.9\%)}\\ 
\hline
H-Optimus-0\cite{hoptimus0} & No & 0.821 & 0.324 & 0.117 \\ 
 & Yes & \textbf{0.835} & \textbf{0.215} \textcolor{mygreen}{(-33.6\%)} & \textbf{0.108} \textcolor{mygreen}{(-7.70\%)}\\ 
\hline
\end{tabular}
\end{table}

\subsection{\textit{ScanGen} with Limited Scanners}
\begin{figure}[t]
\includegraphics[width=\textwidth]{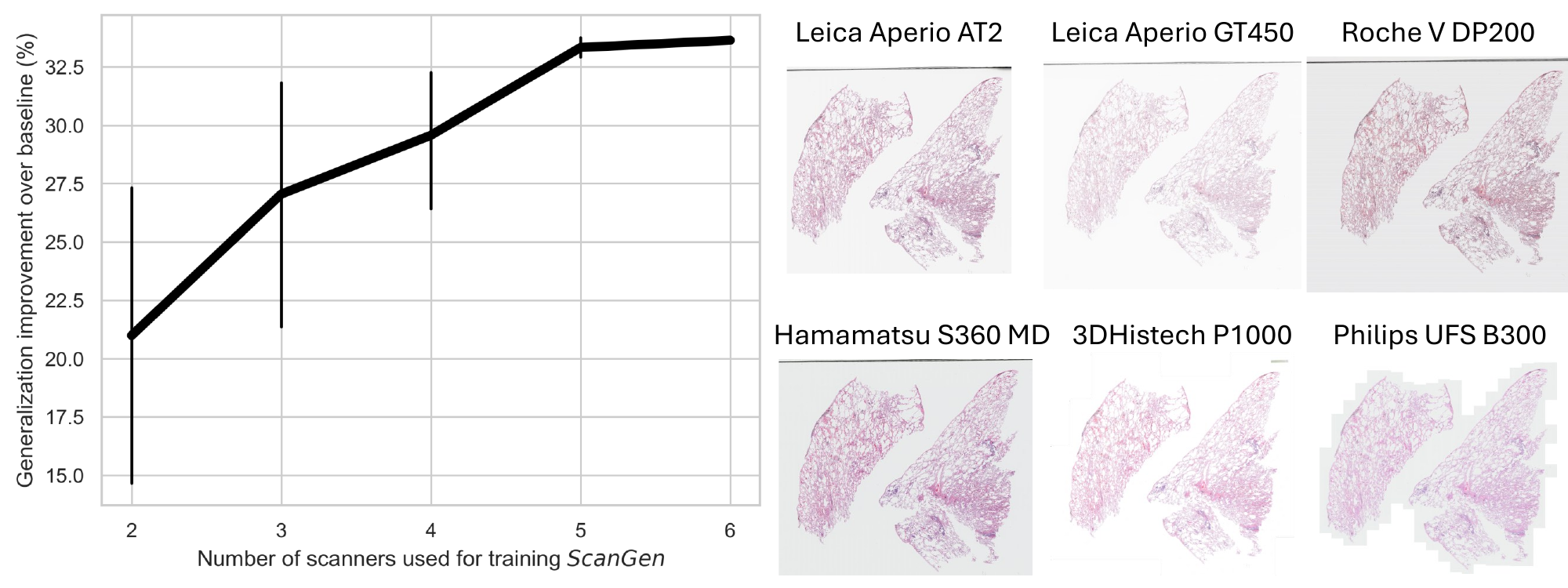}
\caption{
Left: Scanner agreement improves with the number of scanners used for training \textit{ScanGen}. Vertical bars indicate the standard deviation across scanner combinations. Right: Examples of the same specimen digitized using different scanners.
} \label{fig:ablation}
\end{figure}

This section presents an ablation study evaluating the impact of varying the number of scanner types used for \textit{ScanGen} loss during model training. Figure \ref{fig:ablation} highlights the gain of CoV compared to the baseline. Each model is tested on the same external test set containing all six available scanner sources. It examined scenarios where different combinations of two to six scanner types were utilized for the training process.
The plot indicates the monotonic correlation between the number of scanner types and the scanner generalization performance: when more scanner types are used for the \textit{ScanGen} loss, the generalization across scanners increases. Performance improvement converges at five scanners.
Nevertheless, even when only three scanner types are used for training, the scanner generalization on six test scanners significantly improves by $27\%$.

This ablation study also allowed us to investigate which scanner combinations led to the highest or lowest CoV improvement over baseline. We counted how often a certain scanner was present in the best (i.e., lowest CoV) and worst combinations. Doing this for all ablation scenarios from two to five scanners, we found that Leica Aperio GT450 and the Hamamatsu scanners were \textit{always} in the best performing configurations, while the Leica Aperio AT2 and GT450 pair was the one most often associated with the worst performing configurations. This suggests that using a diverse set of scanners for \textit{ScanGen} loss leads to the best generalization, while using similar scanners, like the two Leica Aperio, does not benefit scanner robustness much.

\subsection{Extension to Other MIL Methods}
We also showcase our method's performance on recent MIL techniques like AB-MIL \cite{abmil}, DS-MIL \cite{dsmil}, and SlotMIL \cite{slotmil}. Table \ref{tab:sota} summarizes our findings with H-Optimus-0 features. Our \textit{ScanGen} loss consistently reduced CoV, boosting agreement by $11\%$ to $33\%$, and maintained or surpassed the baseline AUC.

\begin{table}
\centering
    \footnotesize
    \caption{
    Integrating our \textit{ScanGen} loss into other MIL approaches. We report the AUC and scanner CoV values obtained on the held-out test set.}
    \label{tab:sota}
    \begin{tabular}{|l|l|l|l|l|l|l|l|l|}
    \hline
    \multirow{2}{3em}{MIL approach} & \multicolumn{2}{|c|}{GAP} & \multicolumn{2}{|c|}{AB-MIL\cite{abmil}} & \multicolumn{2}{|c|}{DS-MIL\cite{dsmil}} & \multicolumn{2}{|c|}{SlotMIL\cite{slotmil}} \\
    \cline{2-9}
     & AUC & CoV [$\downarrow$] & AUC & CoV [$\downarrow$] & AUC & CoV [$\downarrow$] & AUC & CoV [$\downarrow$] \\
    \hline 
    EGFR  & 0.821 & 0.324 & 0.828 & 0.251 & \textbf{0.834}& 0.265 & 0.822& 0.338\\
    EGFR + \textit{ScanGen} & \textbf{0.835}& \textbf{0.215} & \textbf{0.835}& \textbf{0.224}& 0.828&\textbf{0.214} & \textbf{0.826}& \textbf{0.227}\\
    \hline
    \end{tabular}
\end{table}

\section{Discussion}
This study demonstrates that current FMs for CPath are sensitive to scanners, impacting the model results.
In the global market, scanner distribution is uneven. Leica and Philips lead in America, 3DHistech challenges Leica in Europe, and Hamamatsu is dominant in Asia \cite{pinto2023exploring}. Differences in device performance may affect quality or care between Asia, the U.S., and Europe.
To build real-world CPath tools on a global scale, we need to think about generalization across hospitals in different geographical regions, which requires more robust predictive models. 

Our research is essential in emphasizing a current problem stemming from the swift adoption of FMs in numerous CPath applications.
We first gathered a benchmark comprising six widely used scanners available in the market. Then, we presented empirical evidence of scanner bias across various cutting-edge FMs.
Finally, we offered a solution to mitigate the bias in existing FMs by integrating a scanner generalization loss on top of the standard MIL downstream task.
Ultimately, we aspire that our efforts will motivate future FM developers to address this problem during model training. Nonetheless, training an FM is beyond the reach of most research labs due to the necessity of tens of thousands of samples and significant computational effort. Consequently, our initial focus is on demonstrating the presence of the issue and suggesting a cost-effective solution to enhance the robustness of existing open-source FMs against scanner bias.

    

\begin{credits}

\subsubsection{\discintname}
The authors have no competing interests to declare that are
relevant to the content of this article.
\end{credits}


\clearpage
\bibliographystyle{splncs04}
\bibliography{bibliography}

\begin{thebibliography}{10}
\providecommand{\url}[1]{\texttt{#1}}
\providecommand{\urlprefix}{URL }
\providecommand{\doi}[1]{https://doi.org/#1}

\bibitem{tcga_luad}
Albertina, B., Watson, M., Holback, C., Jarosz, R., Kirk, S., Lee, Y., Rieger-Christ, K., Lemmerman, J.: The cancer genome atlas lung adenocarcinoma collection (tcga-luad)  (2016)

\bibitem{campanella2022h}
Campanella, G., Ho, D., H{\"a}ggstr{\"o}m, I., Becker, A.S., Chang, J., Vanderbilt, C., Fuchs, T.J.: H\&e-based computational biomarker enables universal egfr screening for lung adenocarcinoma. arXiv preprint arXiv:2206.10573  (2022)

\bibitem{uni}
Chen, R.J., Ding, T., Lu, M.Y., Williamson, D.F., Jaume, G., Song, A.H., Chen, B., Zhang, A., Shao, D., Shaban, M., et~al.: Towards a general-purpose foundation model for computational pathology. Nature Medicine  \textbf{30}(3),  850--862 (2024)

\bibitem{CPTAC}
(CPTAC), N.C.I.C.P.T.A.C.: The clinical proteomic tumor analysis consortium lung adenocarcinoma collection (cptac-luad) (version 12) [data set] (2018). \doi{10.7937/K9/TCIA.2018.PAT12TBS}

\bibitem{duenweg2023whole}
Duenweg, S.R., Bobholz, S.A., Lowman, A.K., Stebbins, M.A., Winiarz, A., Nath, B., Kyereme, F., Iczkowski, K.A., LaViolette, P.S.: Whole slide imaging (wsi) scanner differences influence optical and computed properties of digitized prostate cancer histology. Journal of Pathology Informatics  \textbf{14} (2023)

\bibitem{phikon}
Filiot, A., Ghermi, R., Olivier, A., Jacob, P., Fidon, L., Camara, A., Mac~Kain, A., Saillard, C., Schiratti, J.B.: Scaling self-supervised learning for histopathology with masked image modeling. medRxiv  (2023)

\bibitem{hadsell2006dimensionality}
Hadsell, R., Chopra, S., LeCun, Y.: Dimensionality reduction by learning an invariant mapping. In: 2006 IEEE computer society conference on computer vision and pattern recognition (CVPR'06). vol.~2, pp. 1735--1742. IEEE (2006)

\bibitem{abmil}
Ilse, M., Tomczak, J., Welling, M.: Attention-based deep multiple instance learning. In: International conference on machine learning. pp. 2127--2136 (2018)

\bibitem{unrobustfm}
de~Jong, E.D., Marcus, E., Teuwen, J.: Current pathology foundation models are unrobust to medical center differences. arXiv preprint arXiv:2501.18055  (2025)

\bibitem{kang2021stainnet}
Kang, H., Luo, D., Feng, W., Zeng, S., Quan, T., Hu, J., Liu, X.: Stainnet: a fast and robust stain normalization network. Frontiers in Medicine  \textbf{8},  746307 (2021)

\bibitem{kather2020pan}
Kather, J.N., Heij, L.R., Grabsch, H.I., Loeffler, C., Echle, A., Muti, H.S., Krause, J., Niehues, J.M., Sommer, K.A., Bankhead, P., et~al.: Pan-cancer image-based detection of clinically actionable genetic alterations. Nature cancer  \textbf{1}(8),  789--799 (2020)

\bibitem{slotmil}
Keum, S., Kim, S., Lee, S., Lee, J.: Slot-mixup with subsampling: A simple regularization for wsi classification. arXiv preprint arXiv:2311.17466  (2023)

\bibitem{dsmil}
Li, B., Li, Y., Eliceiri, K.W.: Dual-stream multiple instance learning network for whole slide image classification with self-supervised contrastive learning. In: Proceedings of the IEEE/CVF Conference on Computer Vision and Pattern Recognition. pp. 14318--14328 (2021)

\bibitem{macenko2009method}
Macenko, M., Niethammer, M., Marron, J.S., Borland, D., Woosley, J.T., Guan, X., Schmitt, C., Thomas, N.E.: A method for normalizing histology slides for quantitative analysis. In: 2009 IEEE international symposium on biomedical imaging: from nano to macro. pp. 1107--1110. IEEE (2009)

\bibitem{mcinnes2018umap}
McInnes, L., Healy, J., Melville, J.: Umap: Uniform manifold approximation and projection for dimension reduction. arXiv preprint arXiv:1802.03426  (2018)

\bibitem{lunit2022jco}
Park, S., Ock, C.Y., Kim, H., Pereira, S., Park, S., Ma, M., Choi, S., Kim, S., Shin, S., Aum, B.J., et~al.: Artificial intelligence--powered spatial analysis of tumor-infiltrating lymphocytes as complementary biomarker for immune checkpoint inhibition in non--small-cell lung cancer. Journal of Clinical Oncology  \textbf{40}(17),  1916--1928 (2022)

\bibitem{pinto2023exploring}
Pinto, D.G., Bychkov, A., Tsuyama, N., Fukuoka, J., Eloy, C.: Exploring the adoption of digital pathology in clinical settings-insights from a cross-continent study. medRxiv  (2023)

\bibitem{ramanathan2024ensemble}
Ramanathan, V., Pati, P., McNeil, M., Martel, A.L.: Ensemble of prior-guided expert graph models for survival prediction in digital pathology. In: International Conference on Medical Image Computing and Computer-Assisted Intervention. pp. 262--272 (2024)

\bibitem{reinhard2001color}
Reinhard, E., Adhikhmin, M., Gooch, B., Shirley, P.: Color transfer between images. IEEE Computer graphics and applications  \textbf{21}(5),  34--41 (2001)

\bibitem{hoptimus0}
Saillard, C., Jenatton, R., Llinares-López, F., Mariet, Z., Cahané, D., Durand, E., Vert, J.P.: H-optimus-0 (2024), \url{https://github.com/bioptimus/releases/tree/main/models/h-optimus/v0}

\bibitem{shaban2019staingan}
Shaban, M.T., Baur, C., Navab, N., Albarqouni, S.: Staingan: Stain style transfer for digital histological images. In: 2019 Ieee 16th international symposium on biomedical imaging (Isbi 2019). pp. 953--956. IEEE (2019)

\bibitem{teichmann2022end}
Teichmann, M., Aichert, A., Bohnenberger, H., Str{\"o}bel, P., Heimann, T.: End-to-end learning for image-based detection of molecular alterations in digital pathology. In: International Conference on Medical Image Computing and Computer-Assisted Intervention. pp. 88--98 (2022)

\bibitem{gigapath}
Xu, H., Usuyama, N., Bagga, J., Zhang, S., Rao, R., Naumann, T., Wong, C., Gero, Z., Gonz{\'a}lez, J., Gu, Y., et~al.: A whole-slide foundation model for digital pathology from real-world data. Nature pp.~1--8 (2024)

\bibitem{exaonepath}
Yun, J., Hu, Y., Kim, J., Jang, J., Lee, S.: Exaonepath 1.0 patch-level foundation model for pathology. arXiv preprint arXiv:2408.00380  (2024)

\bibitem{zhao2023high}
Zhao, D., Zhao, Y., He, S., Liu, Z., Li, K., Zhang, L., Zhang, X., Wang, S., Che, N., Jin, M.: High accuracy epidermal growth factor receptor mutation prediction via histopathological deep learning. BMC pulmonary medicine  \textbf{23}(1) (2023)

\bibitem{virchow2}
Zimmermann, E., Vorontsov, E., Viret, J., Casson, A., Zelechowski, M., Shaikovski, G., Tenenholtz, N., Hall, J., Klimstra, D., Yousfi, R., et~al.: Virchow2: Scaling self-supervised mixed magnification models in pathology. arXiv preprint arXiv:2408.00738  (2024)

\end{thebibliography}
\end{document}